# Experimental Studies of Compensation of Beam-Beam Effects with Tevatron Electron Lenses

V.Shiltsev[1,5], Yu.Alexahin[1], K.Bishofberger[2], V.Kamerdzhiev[1], V.Parkhomchuk[3], V.Reva[3], N.Solyak[1], D.Wildman[1], X.-L. Zhang[1], F.Zimmermann[4]

[1]*Fermi National Accelerator Laboratory, PO Box 500, Batavia, IL 60510, USA*
[2]*Los Alamos National Laboratory, Los Alamos, NM 87545, USA*
[3]*Budker INP, Novosibirsk, 630090, Russia*
[4]*CERN, European Organization for Nuclear Research, CH-1211 Genève, Switzerland*
[5]*corresponding author, e-mail: shiltsev@fnal.gov*

*Abstract*

Applying the space-charge forces of a low-energy electron beam can lead to a significant improvement of the beam-particle lifetime limit arising from the beam-beam interaction in a high-energy collider [1]. In this article we present the results of various beam experiments with "electron lenses," novel instruments developed for the beam-beam compensation at the Tevatron, which collides 980-GeV proton and antiproton beams. We study the dependencies of the particle betatron tunes on the electron beam current, energy and position; we explore the effects of electron-beam imperfections and noises; and we quantify the improvements of the high-energy beam intensity and the collider luminosity lifetime obtained by the action of the Tevatron Electron Lenses.

TABLE I. Electron Lens and Tevatron Collider parameters.

| Parameter | Symbol | Value | Unit |
|---|---|---|---|
| *Tevatron Electron Lens* | | | |
| e-beam energy (oper./max) | $U_e$ | 5/10 | kV |
| Peak e-current (oper./max) | $J_e$ | 0.6/3 | A |
| Magnetic field in main solenoid | $B_m$ | 30.1 | kG |
| Magnetic field in gun solenoid | $B_g$ | 2.9 | kG |
| e-beam radius in main solenoid | $a_e$ | 2.3 | mm |
| Cathode radius | $a_c$ | 7.5 | mm |
| e-pulse repetition period | $T_0$ | 21 | μs |
| e-pulse width, "0-to-0" | $T_e$ | 0.6 | μs |
| Interaction length | $L_e$ | 2.0 | m |
| *Tevatron Collider* | | | |
| Circumference | $C$ | 6.28 | km |
| Proton(p)/antiproton(a) energy | $E$ | 980 | GeV |
| p- bunch intensity | $N_p$ | 250 | $10^9$ |
| a- bunch intensity (max.) | $N_a$ | 50-100 | $10^9$ |
| Number of bunches | $N_B$ | 36 | |
| Bunch spacing | $T_b$ | 396 | ns |
| p-emittance (normalized, rms) | $\varepsilon_p$ | ≈2.8 | μm |
| a-emittance (normalized, rms) | $\varepsilon_a$ | ≈1.4 | μm |
| Max. initial luminosity/$10^{32}$ | $L_0$ | 2.9 | cm$^{-2}$s$^{-1}$ |
| Beta functions at A11 TEL | $\beta_{y,x}$ | 150/68 | m |
| Beta functions at F48 TEL | $\beta_{y,x}$ | 29/104 | m |
| p-head-on tuneshift (per IP) | $\xi^p$ | 0.010 | |
| a-head-on tuneshift (per IP) | $\xi^a$ | 0.014 | |
| p-long-range tuneshift (max.) | $\Delta Q^p$ | 0.003 | |
| a-long-range tuneshift (max.) | $\Delta Q^a$ | 0.006 | |

# 1. INTRODUCTION

Soon after the introduction of the concept of colliders more than 40 years ago, it was recognized that beam-beam effects due to the immanent electromagnetic interaction of the colliding bunches of charged particles can seriously limit the collider luminosity. The luminosity $L$ of an ultra-relativistic circular collider with equal beam sizes in both transverse directions is equal to

$$L = \gamma f_0 \cdot \frac{N_1 N_2 N_B}{4\pi\varepsilon\beta^*} \cdot H \qquad (1)$$

where $\gamma$ denotes the relativistic gamma factor, $f_0$ the revolution frequency, $N_{1,2}$ the bunch population in both colliding beams (distinguished by the subscripts '1' and '2'), $N_B$ the total number of bunches, $\varepsilon$ the normalized emittance which is related to the transverse rms beam size $\sigma$ at the interaction point (IP) through the beta function (Twiss parameter of the optical focusing latter) at the IP, $\beta^*$, via $\varepsilon = \gamma\sigma^2/\beta^*$, and, finally, $H$ a geometric reduction factor (less than unity) due to a so-called "hour-glass" effect (variation of the beta function over the length of the collision region) and possibly also due to incomplete geometric overlap of the beams at the IP. The rate of high-energy physics (HEP) reactions is equal to product of the luminosity and the reaction cross section. Strong transverse electric and magnetic (EM) fields of the opposite bunches can lead to a blowup of the beam emittances, a significant loss of the beam intensity and an unacceptable background in the HEP detectors. The fields of a bunch with round Gaussian charge distribution of rms size $\sigma$ at radius $r$ scale as:

$$(E_r, B_\theta) \propto \frac{2eN_2}{r} \cdot \left(1 - \exp(-r^2/2\sigma^2)\right) \qquad (2).$$

The commonly used figure of merit for such EM interaction is the *beam-beam parameter* $\xi \equiv r_0 N/(4\pi\varepsilon)$, where $r_0 = e^2/mc^2$ is the classical particle radius (in cgs units) and $N_2$ the number of particles in the opposing bunch [2]. This parameter is approximately equal to the shift of the betatron tune $Q = f_\beta/f_0$ of core particles due to the beam-beam forces. The tune $Q_{x,y}$, a key stability parameter, is the number of periods of a particle's horizontal or vertical oscillations in the focusing lattice over one turn around the ring. While core particles undergo a significant tune shift, the tune shift is negligible for halo particles with large oscillation amplitudes. The EM forces drive nonlinear resonances $n = kQ_x + lQ_y$ (with $n,k,l$ denoting integers) which can result in an instability of the particle motion and ultimately particle loss. The beam-beam limit in modern hadron colliders is empirically found at $\xi^{max} \cdot N_{IP} \approx 0.01 - 0.025$ ($N_{IP}$ is the number of IPs), while it can exceed $\xi^{max} \cdot N_{IP} \approx 0.1$ in high energy $e+e-$ colliders [3].

In the Tevatron collider, beams of protons and antiprotons share the same beam pipe and magnet aperture and, in order to reduce the number of head-on interactions, the beams are placed on separated orbits everywhere except the CDF and D0 detectors IPs by using electrostatic separators. The long-range interaction between separated beams can be derived from Eq.(2) if $r \gg \sigma$. The effect scales as $(E,B) \sim N_2/r$. Besides being nonlinear, the long-range effects vary from bunch to bunch which makes them hard to treat. Head-on and long-range beam-beam effects seriously affect the collider performance, reducing the luminosity integral per HEP store (period of continuous collisions) by 10-30%. Tuning the collider operation for optimal performance becomes more and more cumbersome as the beam intensities and luminosity become larger. A comprehensive review of the beam-beam effects in the Tevatron Collider Run II can be found in Ref. [4].

Electron lenses were proposed for compensation of both long-range and head-on beam-beam effects in the Fermilab's Tevatron [5]. An electron lens employs a low-energy beam of electrons that collides with the high-energy bunches over an extended length $L_e$. Electron space-charge forces are linear at distances smaller than the characteristic beam radius $r < a_e$ (with $a_e$ signifying the electron-beam radius), but scale as $1/r$ for $r > a_e$, similar to Eq.(2). Correspondingly, such a lens can be used for linear and nonlinear force compensation by manipulating the beam-size ratio $a_e/\sigma$ and the current-density distribution $j_e(r)$. The electron current profile (and thus the EM field profiles) can easily be changed for different applications. Moreover, the electron-beam current can be adjusted between individual bunches, equalizing the bunch-to-bunch differences and optimizing the performance of all bunches in a multi-bunch collider. Another advantage of the electron lens is the absence of nuclear interactions with (anti)proton beams since the electron beam acts only through EM forces – so, there are no radiation issues. Finally, coherent instabilities are greatly suppressed because of two reasons: a) fresh electrons interact with the high-energy particles on each turn, eliminating the possibility for multi-turn amplification due to a memory of the electron beam; b) the electron beam is made very rigid transversely by immersion in a very strong magnetic field. (This field also helps to keep the electron beam straight and its distribution unaffected by the high-energy beam EM fields). For effective beam-beam compensation (BBC), the number of electrons required is roughly the same as the number of particles in the high-energy bunch. This implies that most modern colliders only need a few amperes of electron-beam current to achieve successful BBC.

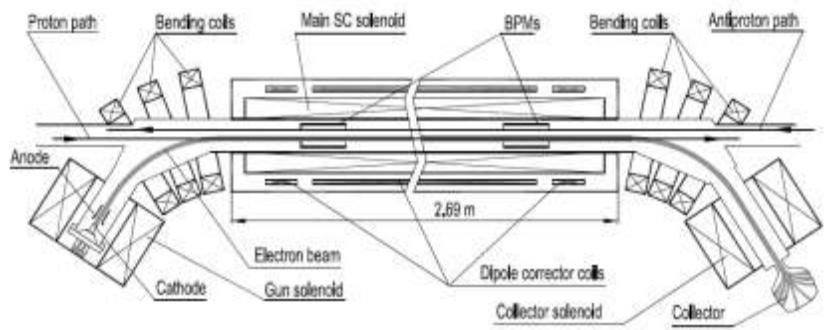

Figure 1: General layout of the Tevatron Electron Lens and its main components.

Two Tevatron Electron Lenses (TELs) were built and installed in two different locations of the Tevatron ring, A11 and F48. Figure 1 depicts a general layout of the TEL. A list of the TEL and relevant Tevatron parameters is given in Table 1. Three conditions were empirically found to be crucial for successful BBC: 1) the electron beam must be transversely centered on the proton (or on antiproton) bunches, within 0.2–0.5 mm, along the entire interaction length; b) fluctuations in the current needs to be less than one percent, and timing within one nanosecond, in order to minimize emittance growth; and c) the transverse profile of the current density should have no sharp edges.

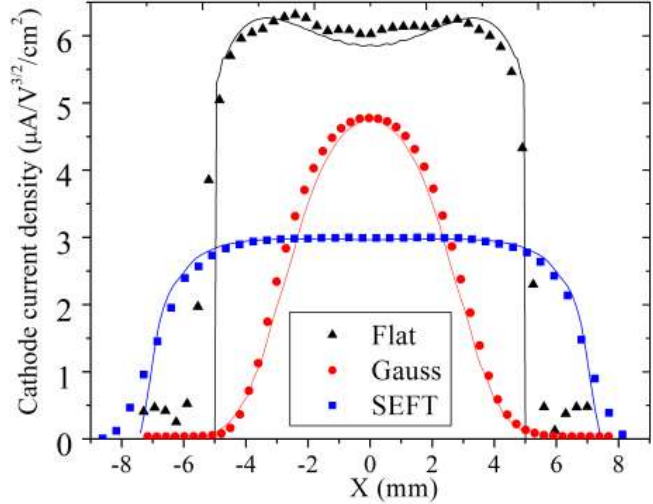

Figure 2: Three profiles of the electron current density at the electron gun cathode: black – flat-top profile; red – Gaussian profile; blue – smooth-edged, flat-top (SEFT) profile. Symbols represent the measured data, the solid lines are simulation results. All data refer to an anode-cathode voltage of 10 kV.

The shape of the electron current density distribution is determined by the electron gun (geometry of its electrodes and voltages). In our beam studies, we employed three types of electron guns generating flat-top, Gaussian, and smoothed-edge-flat-top distributions, as shown in Fig.2, and for all experimental results presented in this report we mention which gun was being used at the time. The desired distributions were generated from a convex impregnated tungsten thermo-cathode of either 5mm or 7.5-mm radius [6].

In order to keep the electron beam straight and its distribution unaffected by its own space-charge as well as from the high-energy beam's EM fields, the electron beam is immersed in a strong magnetic field of about $B_{gun}$=3 kG (4 kG maximum) at the electron-gun cathode and some $B_{gun}$ =30 kG (65 kG maximum) inside the main superconducting (SC) solenoid. The TEL magnetic system compresses the electron-beam cross-section area in the interaction region by the factor of $B_{main}/B_{gun} \approx 10$ (variable from 2 to 30), proportionally increasing the current density of the electron beam in the interaction region. Most experiments have been performed using an electron beam with an energy between 5 keV and 10 keV and a current between 0.5 A and 3 A. The deviations of the magnetic field lines from a straight line are less than ±100 μm over the entire length of the SC solenoid. The strongly magnetized electron beam follows these field lines; therefore, it does not deviate from the straight Tevatron beam trajectory by more than 20% of the Tevatron beam rms size $\sigma \approx 0.5 - 0.7 \text{ mm}$ in the location of the TELs. To assure proper placement and overlap of the electron beam with the high-energy beam all along the interaction length, beam position monitors (BPMs) and steering magnets are employed in the TELs. Two pairs of BPMs, one horizontal and one vertical, are located at either end of the main solenoid. The BPMs simultaneously report measured transverse positions for all three beams passing through (electrons, protons, and antiprotons), thus allowing the electron beam to be centered either on the antiproton beam or on the proton one. The BPM accuracy is about 0.1 mm. The electron beam steering is done by adjusting currents in superconducting dipole correctors (2 – 8 kG maximum field) installed inside the main solenoid cryostat. Figure 3 schematically presents all three beams inside the A11 TEL beam tube. It indicates the relative sizes of 1σ, 2σ, and 3σ ellipses for the 980-GeV protons and antiprotons as well as for the round (SEFT) electron beam centered on the proton beam.

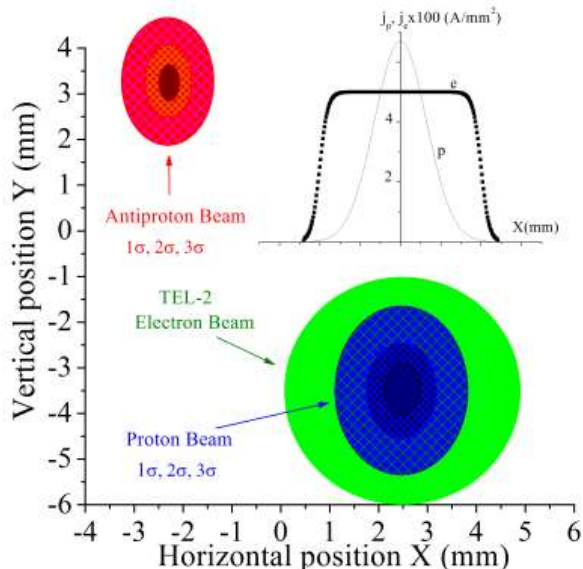

Figure 3: Schematic of the transverse electron beam alignment with respect to the proton and antiproton beams.

In order to enable operation on a single bunch in the Tevatron which has a bunch spacing of 396 ns, the anode voltage, and consequently the beam current, are modulated with a characteristic on-off time of about 0.6 μs and a repetition rate equal to the Tevatron revolution frequency of $f_0 = 47.7$ kHz by using a HV Marx pulse generator or a HV RF tube base amplifier. Electron currents leaving the cathode, into the collector, and onto the collector entrance electrode are measured by three inductive coils. The vacuum in the TEL is usually under $2 \cdot 10^{-9}$ Torr, sustained by three ion pumps with 300 l/s total pumping speed. The TEL magnets have a minimal effect on the 980-GeV proton-beam path - with the orbit distortion around the ring staying within ±0.2 mm. The broadband impedance of the TEL components is $|Z/n| < 0.1\,\Omega$, much less than the total Tevatron impedance of $5 \pm 3\,\Omega$. More detailed descriptions of the TEL, its components and the associated beam diagnostics can be found in Refs. [7] and [8] and references therein.

The successful use of electrons for the compensation of beam-beam effects in the Tevatron – namely, the improvement of the colliding beam lifetime by the TELs – was already reported in Ref. [1]. In this article we describe in full detail a variety of beam experiments performed with the electron lenses. In the next section we discuss the experimentally measured dependencies of the particle betatron tunes on electron beam current, energy and position. Section 3 describes the effects arising from imperfections and noises of the electron beam, as well as observations of longitudinal waves excited in the electron beam. The results presented in Section 4 focus on the effect of the electron current-density profile on the Tevatron beam losses. The reduction of the antiproton emittance-growth rate using the TELs is discussed in Section 5. Finally, Section 6 presents the observed improvements in the lifetime of the Tevatron beam intensity and in the Tevatron luminosity under the action of the Tevatron Electron Lenses.

## 2. SHIFT OF BETATRON TUNES BY ELECTRON LENSES

A perfectly steered round electron beam with current density distribution $j_e(r)$, will shift the betatron tunes $Q_{x,y}$ of small amplitude high-energy (anti-)protons by [5]:

$$dQ_{x,y} = \pm \frac{\beta_{x,y} L_e r_p}{2\gamma ec} \cdot j_e \cdot \left(\frac{1 \mp \beta_e}{\beta_e}\right) \quad (3),$$

where the sign reflects focusing for protons and defocusing for antiprotons, $\beta_e = v_e/c$ is the electron beam velocity, $\beta_{x,y}$ are the beta-functions at the location of the lens, $L_e$ denotes the effective interaction length between the electron beam and the protons or antiprotons, $r_p = e^2/mc^2 = 1.53 \cdot 10^{-18}$ m is the classical proton radius, and $\gamma_p = 1044$ the relativistic Lorentz factor for 980-GeV protons. If the electron beam is much wider than the (anti)proton beam, then all of the high-energy particles acquire the same $dQ_{x,y}$. The factor $1 \pm \beta_e$ reflects the fact that the contribution of the magnetic force is $\beta_e$ times the electric force contribution, and its sign depends on the direction of the electron beam; both TELs direct the beam against the antiproton flow. The first TEL, "TEL-1," is installed in the F48 sector, where the horizontal beta-function $\beta_x = 104$ m is larger than the vertical beta-function $\beta_y = 29$ m. Correspondingly, it mainly affects horizontal beam tunes. "TEL-2," the second TEL, is placed in the A11 sector, where $\beta_y = 150$ m is larger than $\beta_x = 68$ m, so that it affects the vertical tune more strongly. Note that the electron beams are round in both lenses. The shifts of the Tevatron betatron tunes by the TELs were measured in great detail and found to be in very good agreement with Eq.(3).

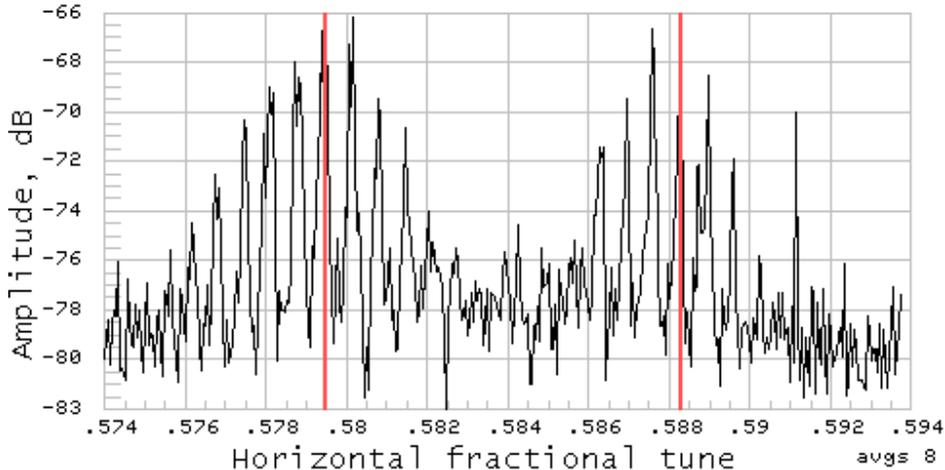

Figure 4: Tevatron horizontal 21-MHz Schottky spectrum of three 980 GeV proton bunches with one bunch tune-shifted by TEL1 (electron current $J_e = 2.6$A, electron energy $U_c = 7.6$ kV, flat-top electron gun).

21 MHz and 1.7 GHz Schottky detectors are used in Tevatron to measure the tunes of the proton and antiproton bunches [9]. The 21 MHz vertical and horizontal Schottky detectors have very good tune resolution but are not directional and do not allow to measure tunes of individual bunches. The 1.7 GHz detectors are less accurate but they have larger bandwidth, allow to perform measurements for individual bunches and are directional, so, proton and antiproton signals are measured separately. Figure 4 depicts the 21 MHz Schottky spectra during one of the TEL-1 studies: only three proton bunches were circulating in the Tevatron (without antiprotons), and the electron pulse was timed on only one of the three bunches. A series of synchro-betatron sidebands, separated by the synchrotron tune $Q_s \approx 0.0007$, on the left correspond to the signal from the two bunches not affected by the TEL-1, and their central line (highlighted by a marker) is found at the fractional horizontal tune of $Q_x = 0.5795$. A similar series of lines on the right, which belong to the TEL-affected proton bunch, is shifted by $dQ_x=0.0082$ to $Q_x=0.5877$. The shape of the Schottky spectra depends on the proton intensity, machine chromaticity, tuning, working point, etc. Application of the electron beam may or may not cause a variation of the spectral shape. The typical tune measurement error for the 21 MHz detector is estimated to be about $\delta Q \approx \pm 0.0002$.

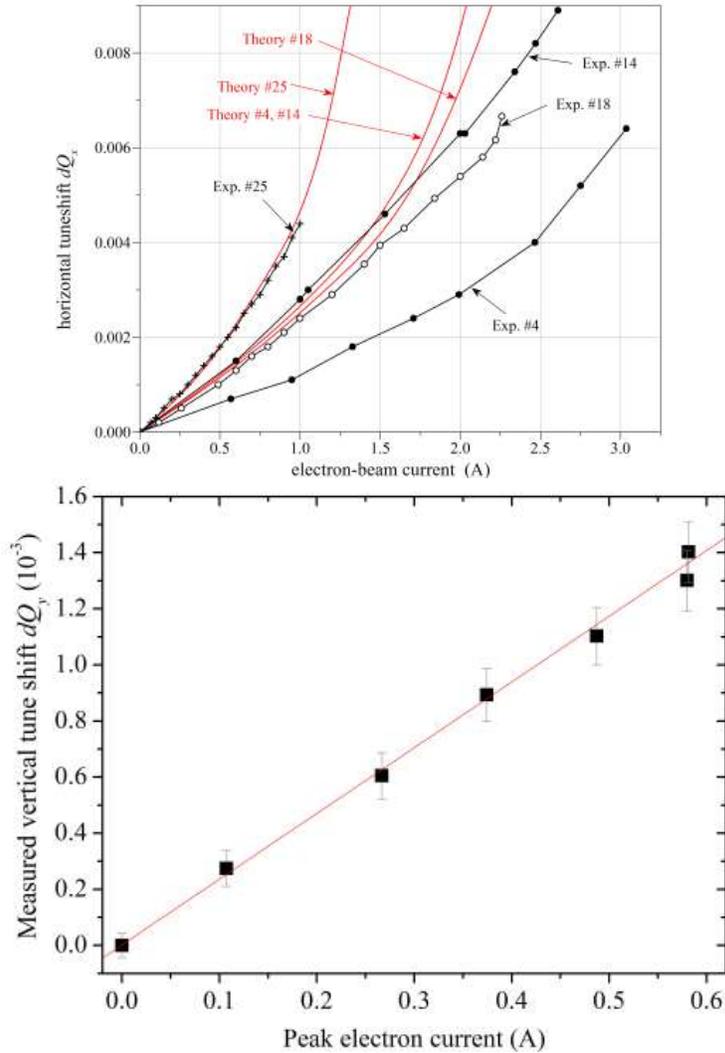

Figure 5 : Shift of the 980-GeV proton betatron tunes by electron lenses : a) top – horizontal tune shift vs. the electron current in TEL-1, over several separate experiments (flat-top beam profile) ; b) bottom – vertical tune shift vs. TEL-2 current (SEFT gun, store #4224). Solid lines are theoretical predictions according to Eq.(3).

After properly synchronizing the electron beam for maximum effect, we have studied the dependence of $dQ_x$ on the peak electron current. The results are presented in Fig.5 and compared with Eq.(3). The theoretical dependence is nonlinear because the electron energy inside the vacuum pipe (and thus $\beta_e$) decreases with increasing current due to the electron space charge, $U_{e.}=U_{c.}-gQ_{SC}$, where $g$ is a factor depending on the chamber and beam geometries. In Figure 5a, the tuneshift data measured by 21 MHz Schottky detector during four selected experiments with TEL-1 is shown. The TEL cathode voltage was set to -7.5 kV in experiments #4 and #14, to -8kV in the experiment #18 and to -4.7kV in the experiment #25. The voltage difference is the reason of significant variation of slopes in Fig.5a. The experiment labeled as Experiment #4 is in fact one of the earliest with the TEL-1 after it was installed in the Tevatron and the large discrepancy between measured tuneshift and theoretical prediction was due to poor translation alignment (see also discussion below). In the other three experiments with better alignment, the maximum discrepancy did not exceed ~20% at $J_{e.}=2$ A. There are systematic errors in a number of the parameters used in numerically calculating Eq.(3). For example, $a_e^2$ is known only to within ±10%, the effective length $L_e$ depends on the precision of the steering and may vary within ±10%, and the electron current calibration contributes about ±5% error [10]. Figure 5b presents the vertical tune shift induced by the TEL-2 electron current from the SEFT gun. There is an excellent agreement between the tune shift measured by the 1.7 GHz Schottky tune monitor and the theory. The dependence of the tune shift on the electron energy also agrees with the theoretical predictions; pertinent results displayed in Fig.6 show 980-GeV antiproton tune-shift measurements at various cathode voltages $U_c$, ranging from -6 kV to -13 kV. As the total electron beam current (which is determined by the gun cathode-anode voltage difference, and shown by the dashed line) was kept constant, the total electron space-charge $Q_{SC}$ grew for smaller values of $U_c$, inducing correspondingly larger tune shift.

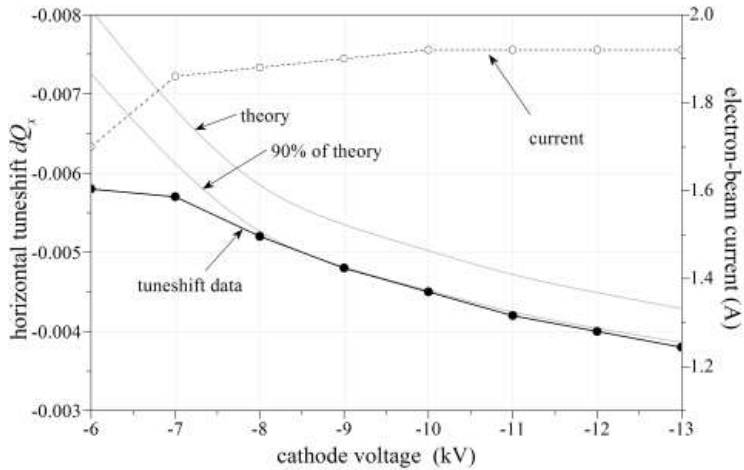

Figure 6 : Horizontal tune shift of 980-GeV antiprotons versus cathode voltage (electron energy). This data used TEL-1 with the flat-top electron gun.

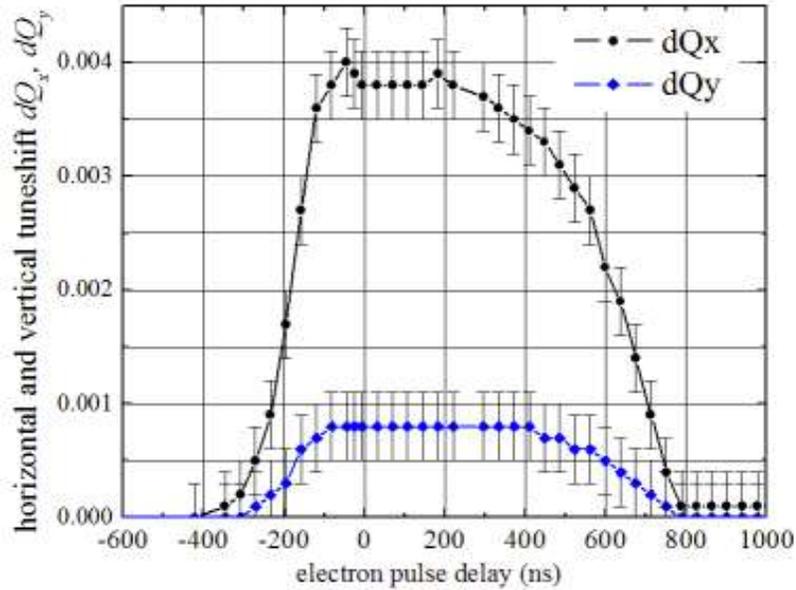

Figure 7 : TEL-1-induced shift of 980 GeV proton horizontal (black line) and vertical (blue line) betatron tune versus delay time for an 800-ns long pulse of electrons (1.96 A peak current, -6.0 kV cathode voltage, flat-top gun).

Fig.7 illustrates how the proton tune shifts depend on the time delay between the 2A electron pulse and the arrival of the proton bunch. With a small correction for the electron beam propagation time along the TEL interaction region (~50ns), the tune shift follows the electron pulse shape. One important conclusion is that the electron pulse is short enough to allow shifting the tune of any given bunch without touching its neighbors 396-ns away. This feature was extremely useful over the entire series of beam studies with the TELs, as it allowed us to vary electron beam parameters and to tune up the lens affecting just one bunch out of 36 circulating in the machine. As seen in Fig.7, the horizontal tune shift is about four times the vertical one $dQ_x/dQ_y =0.0037/0.0008=4.6$, which is close to the design beta function ratio $\beta_x/\beta_y=101/28=3.6$. The remaining discrepancy can be explained by the beta-function measurement error, which could be as big as ±20%, or a small ellipticity of the electron beam, or a mis-steering of the electron beam, which may play a role if it is not small compared with $a_e$.

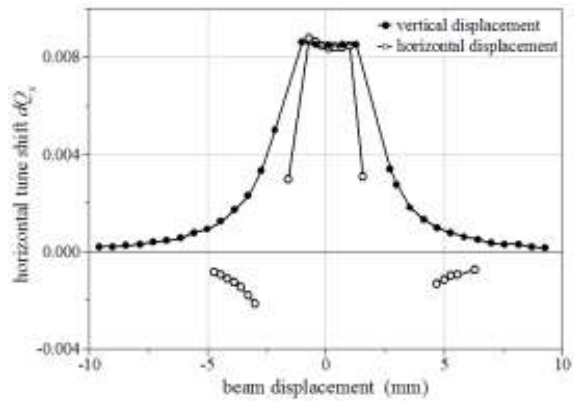

Figure 8 : TEL1-induced shift of 980 GeV proton horizontal betatron tune versus vertical (filled circles) and horizontal (open circles) electron beam displacement ($J_e$ =1A, $U_c$ =6.0 kV, flat-top electron gun).

As long as the proton beam travels inside a wider electron beam, the proton tune shift does not depend much on the electron beam position $d_x, d_y$; for example, in the case of a 1-A electron beam, $dQ_x(d_x,d_y) \approx dQ_{max}$ = 0.0021 if the displacement $|d_{x,y}|$ < 2 mm, as illustrated in Fig.8. However, when the distance between the centers of the two beams exceeds the electron-beam radius $a_e$ then one expects $dQ_x(d_x,d_y=0) \approx -dQ_{max}/(d_x/a_e)^2$, $|d_x|>a_e$, and $dQ_x(d_x=0,d_y) \approx +dQ_{max}/(d_y/a_e)^2$ $|d_y|>a_e$. Such a change of the sign of the tuneshift is clearly seen in Fig.8.

To summarize, the experimentally observed shifts of the betatron tunes of 980 GeV protons and antiprotons due to TELs agree reasonably well with theoretical predictions.

## 3. STUDIES OF ELECTRON BEAM FLUCTUATIONS EFFECTS

Fluctuations in the electron beam can lead to several phenomena in the high energy beams: a) turn-by-turn electron current jitter and transverse electron beam position fluctuations can blow up the transverse emittance – e.g. theory [5] predicts sizable growth if the rms current fluctuations $\delta J$ exceed (3-10)mA or if the position jitter $\delta X$ in a multi-Ampere beam is bigger than 0.2μm; b) electron pulse timing jitter results in similar effects if the pulse does not have a flat-top; c) low-frequency variations of the parameters may result in the orbit or/and tune variations leading to a faster dynamical diffusion. High-frequency current fluctuations measured directly from the TEL-1 BPM signals using a 15 bit ADC segmented memory scope showed $(\delta J/J) \sim (4\text{-}10) \cdot 10^{-4}$ for pulses of current $J \approx 0.3\text{-}0.5A$; we also estimate an upper limit on the beam position stability of about 10 μm [11].

To observe the effect that the level of fluctuations has on the antiproton emittance growth, the electron gun HV modulator pulse circuit was modified to produce a random-amplitude pulse. This was established by setting an average pulse amplitude modulated by a noise generator. At different noise levels, the 980 GeV antiproton bunch emittance is observed long enough to record its growth by so called "Flying Wires" beam size monitors [9]. Figure 9 shows that the emittance growth increases – as expected in [5] - with the square of the amplitude fluctuations. The Tevatron high-intensity proton and antiproton beams, without the TEL, have a typical emittance growth of 0.04-0.2 πmm-mrad/hr. If the TEL is allowed to only enhance the emittance growth by 0.01 mm-mrad/hr, added in quadrature to the Tevatron's inherent emittance growth, then according to the measured dependence in Fig. 9 this limit corresponds to about 3mA peak-to-peak current variation.

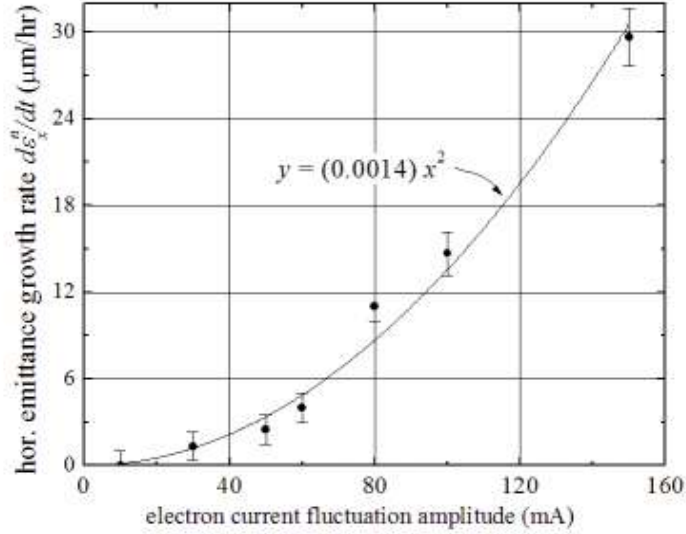

Figure 9 : Horizontal emittance growth rate of 980 GeV antiprotons vs TEL 1 electron current fluctuation amplitude (flat-top electron gun).

Another source of fluctuations is timing jitter. It was noted that a large timing jitter of about 10ns peak-to-peak (due to an instability of the synchronization electronics) leads to a detectable emittance growth and to a significant increase in the 21 MHz Schottky detector signal power. This effect was particularly large on the rising and falling slopes of the electron pulse (where the derivative of the electron current $dJ_e/dt$ is large). Elimination of the source of the instability and use of optical cables for synchronization of the TEL pulsing with respect to the Tevatron RF allowed us to reduce the jitter to less than 1 ns and to bring the corresponding emittance growth to within the tolerable level. In nominal operation conditions of the TEL, without the noise generator, with low-frequency current variations reduced to under 5mA, and with timing jitter under control, we observed no detectable additional emittance growth within the resolution of our beam size monitors $(d\varepsilon/dt)_{rms}$~0.02 $\pi$ μm/hr.

Monitoring the 21 MHz Schottky power is useful for studying the effect of the beam displacement. Figure 10 presents the dependence of the Schottky power on the TEL2 vertical beam position (the electron beam is perfectly aligned with the proton beam horizontally). An additional electron current noise of about 50mA peak-to-peak was induced in order to make the effect more prominent. The study was performed at the end of HEP store #5152 with both proton and antiproton beams present. One can see that the Schottky power rises with increasing separation of the electron and proton beams – approximately as $P \sim E_y^2$, where the electric field due to the electron space charge is given by Eq.(2). The asymmetry of the measured power with respect to vertical position can be explained by the effect of the TEL2 on the antiproton beam – its position is indicated by the red ellipse in Fig. 10; see also Fig. 3 for reference (note that the 21 MHz Schottky monitor is not directional and reports the power from both the proton and antiproton beams circulating in the machine).

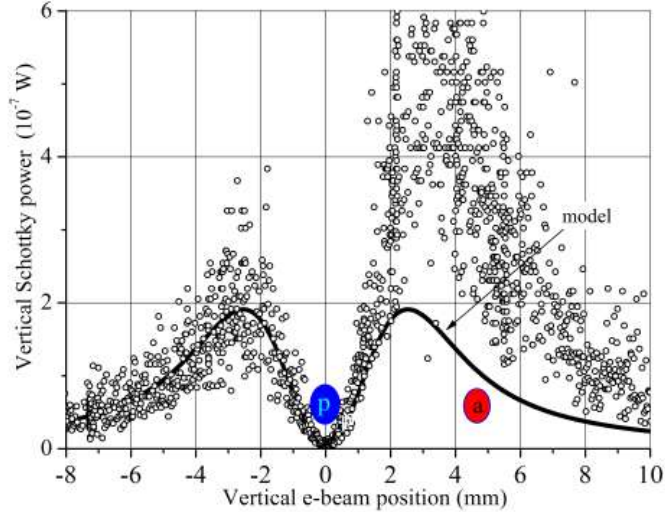
Figure 10: Vertical Schottky power vs TEL2 electron beam vertical in store #5152. Electron current noise amplitude about $\delta J_e$=50mA peak-to-peak (SEFT electron gun).

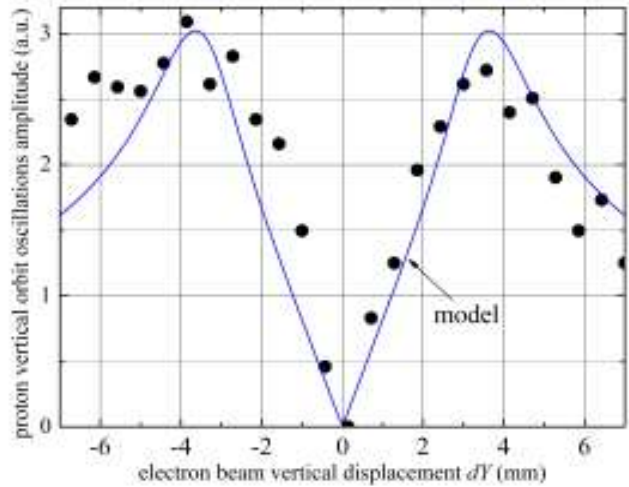
Figure 11 : $K$-modulation position scan with TEL1 (flat-top electron gun).

"Tickling" of the proton orbit by the electron beam can be used for electron beam steering. The idea is similar to the "$K$-modulation" in the beam based alignment [12]: variation of the electron current in the electron lens causes variations in the proton beam orbit around the ring if the electron lens beam is not centered. Figure 11 shows the rms amplitude of the vertical proton orbit variation at the Tevatron BPM located in the A0 sector vs. the vertical displacement of the TEL1 electron beam at F48. The current of the latter was modulated as $J_e$ [A]=1.02+ 0.18sin($2\pi$*107Hz). The amplitude becomes equal to 0 if the proton beam goes through the center of the electron beam. Maximum amplitude of the orbit response at 107 Hz is about a few micrometers. The 7-mm distance between the two peaks reflects an effective diameter of the electron current distribution, and, thus, indicates some angular misalignment of the electron beam because it exceeds the electron beam diameter $2a_e \approx$3.5 mm. Therefore, steering by the orbit tickling should concentrate not only on the search of the minimum orbit response, but also on getting the two maxima closer to each other.

Yet another indication of a good steering of the electron beam is the observation of longitudinal space-charge waves in the electron beam induced by the proton bunches. Figure 12 presents digital scope records of the TEL2 BPM signal (pickup) and electron current pulses measured at the cathode and at the collector of the lens. The difference between the last two is that the electron current pulse in the collector exhibits additional waves (wiggles starting around $t=0$ ns) due to interaction with the protons. The amplitude of the waves is about 5% of the maximum total electron current. Any significant separation of the electrons and protons (several mm transversely, or timing the electron pulse away from the proton bunches, e.g. in the abort gap) leads to the disappearance of the waves and to the collector signal becoming like the one at the gun. Detailed theoretical analyses and extensive experimental studies of these waves have been reported in [13].

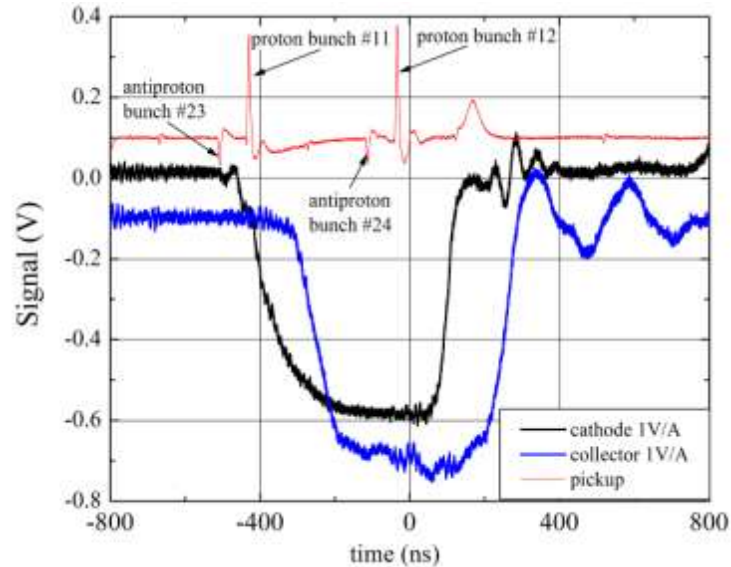

Figure 12: Longitudinal waves in the TEL2 electron beam excited by the interaction with a proton bunch: red – TEL2 beam pickup signal, black – electron current measured at the SEFT electron gun cathode, blue – electron current measured at the collector.

## 4. EFFECT OF ELECTRON BEAM CURRENT PROFILE ON HIGH ENERGY BEAM LIFETIME

Typically the lifetime of (anti)proton bunches in the Tevatron, before collisions and without the TEL, is on the order of 150-600 hours. As a bunch traverses the Tevatron ring billions of times, several mechanisms contribute to a gradual growth in its emittance of about 0.04-0.2π mm-mrad/hr. These include: residual-gas scattering, intra-beam scattering, and fluctuations in the ring elements [14]. As the bunch size increases, particles gradually diffuse to larger oscillation amplitudes until they finally collide with some aperture restriction, usually one of many retractable collimators inserted in the Tevatron beam pipe. As shown in Ref. [4] and as will be discussed further below, the beam-beam interaction with either the electron beam or the opposite high-energy beam can lead to a significant decrease of the beam lifetime.

Originally, it was planned to generate an electron beam wide enough to cover all of the high energy proton or antiproton beams – and this large size was thought to be helpful to maintain low particles loss rates. In reality, however, there are always particles with amplitudes beyond the electron beam cross section. For such particles with oscillations larger than the size of the electron beam, the electric field due to the electron space charge is no longer linear with the transverse displacement and the resulting nonlinearities may significantly change the particle dynamics depending on the electron current distribution. As we found experimentally, in the worst case of the flat-top electron beam, the electron beam edges act as a ``gentle'' collimator, since the outlying particles are slowly driven out of the bunch until they eventually hit the collimators.

A convenient way to measure this effect is to observe the bunch size as the TEL trims away extraneous particles. In Fig.13, one bunch was monitored over hundred minutes as the TEL-1 was "shaving" the bunch size. The current of the TEL was initially set to 1 A for the first 45 minutes. After a ten-minute respite, the current was increased to 2 A (these settings are shown above the plot). After about 85 minutes, the TEL-1 was purposefully mis-steered in order to observe a ``blowup'' in the bunch sizes. The upper data in Fig.13 show the horizontal and vertical beam sizes measured many times during this process. Also indicated is the longitudinal bunch size.

The open circles show the intensity of the bunch during this process. One can see a fast initial decreasing of sizes, but after about ten minutes, the rate of decrease drops significantly; this implies that the large-amplitude particles have been removed, and the core is more stable inside the electron beam. In addition, the increase of the TEL-1 current to 2 A was expected to worsen the bunch-size lifetime, but the smaller bunch was well preserved for the remaining time that the TEL-1 electron beam was on and centered on the proton beam. The stability of the bunch size is remarkable, suggesting that the flattop profile was ideal for the small bunch size.

The bunch intensity decay rate also decreases significantly after a short interval of faster losses, and when the electron current is doubled, the decay rate is nearly unchanged. After the bunch was observed for a while, the electron beam was moved transversely so that the bunch intercepted the edge of the electron beam. As expected, the particles were suddenly experiencing extremely nonlinear forces, causing emittance (and size) growth, shown by the bump in the upper plot of Fig.13, and heavy losses, shown by the fast decline of the lower plot.

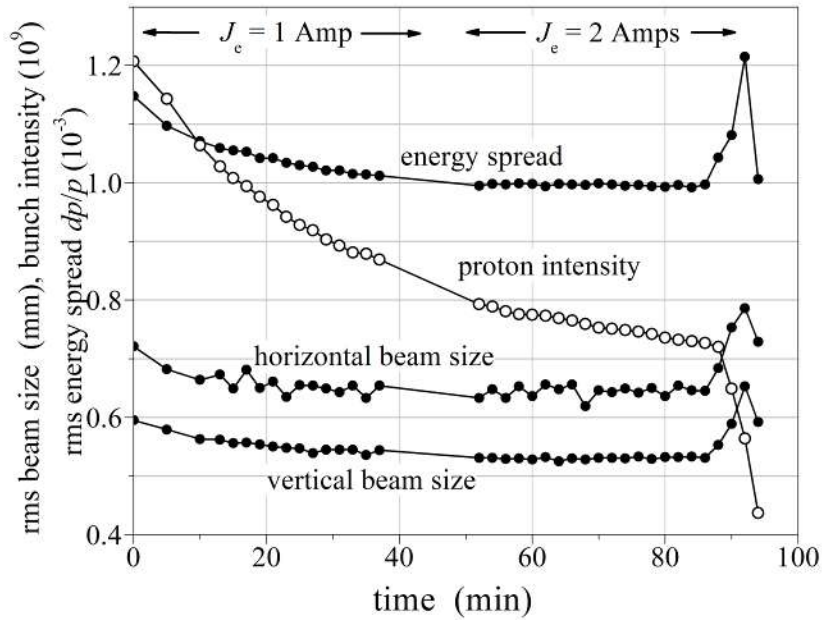

Fig.13: Scraping of a proton bunch due to interaction with the TEL-1 electron beam (flat-top electron current distribution).

Fig.14 presents the results of an experiment where the proton loss rate, as indicated by the CDF detector beam-loss monitors, was measured as the TEL current was changed. This test was performed with the flat-top electron beam centered on a single proton bunch, and the cathode voltage was -10 kV. The losses varied from about 250 Hz at low currents to 1 kHz at the highest currents. At zero current, the average loss rate was approximately 230 Hz over a large portion of an hour. The losses data were converted into lifetimes, producing more tangible results. This conversion was straightforward, since the lifetime is given by $\tau = -kN/(dN/dt)$, where $N$ was the current total number of particles and $(dN/dt)$ was the loss rate measured by the beam-loss monitors. The constant $k$ was determined from a calibration test. During a period of a couple hours, the bunch was allowed to proceed without any changes made to the TEL-1 After this amount of time, the number of particles had diminished enough to directly compute the lifetime. Since the number of particles and the average loss rate were also known, the constant $k$ could be derived for the conditions of the experiment [7].

The maximum current in this experiment was about $J_e=0.75$A and the corresponding proton tune shift was about $dQ_x = 0.0022$. Despite of the small tune shift, the proton bunch lifetime at the higher electron-beam currents was less than 50 hours, significantly less than the typical 175-hour lifetime without interference. While it was impossible to guarantee that the electron beam was perfectly centered on the proton orbit, adjustments of the beam position yielded no improvement in the bunch lifetime. The solid line in Fig.14 represents the fit $\tau^{-1}$ $[1/hr]=1/150+J_e^{2}/30$. In this experiment, the electron radius was $a_e =1.6$ mm, and the proton rms beam size at the location of the TEL-1 was about $\sigma_x =0.8$ mm, corresponding to an rms normalized emittance of about 5 $\pi$mm-mrad.

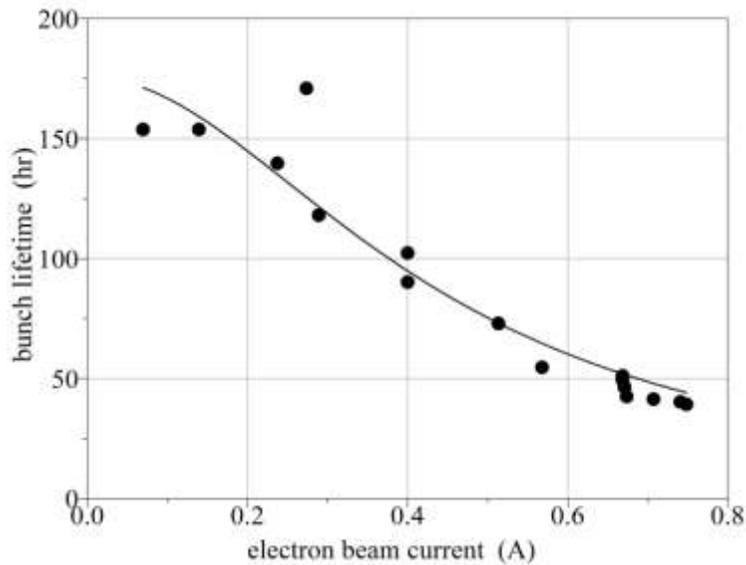

Figure 14 : Dependence of the proton bunch intensity lifetime on the TEL-1 current (flat top electron current distribution).

The examples presented above and the unsatisfactory low beam lifetime in the BBC experiments during HEP stores, convinced us that the flat-top electron current distribution edges introduce severe constraints on the performance of the TEL-1. A Gaussian gun was designed to obtain much smoother edges, so that particles at large betatron amplitudes would not feel strongly nonlinear space-charge forces. Figure 2 compares the current density profiles for the flat-top and Gaussian gun. In order to quantify the differences between these two guns, a scan of working points ($Q_x$, $Q_y$) was performed with each of them. In this test, the Tevatron horizontal and vertical tunes were independently adjusted to cover approximately a 0.020 span in both dimensions. By adjusting the tunes in 0.002 increments, the loss rate was measured, recorded, converted to a lifetime, and plotted in Fig.15.

In order to simplify the interpretation of the results, both guns were set to currents such that the horizontal tuneshift was 0.004 and the vertical was 0.0013. The Tevatron is equipped with tune-adjustment quadrupole magnet circuits, which provided a convenient way to adjust the tune. Confirmation of the correct tune was possible with the Schottky detectors, but sometimes when the loss rate was high, an accurate measurement of the tune was difficult to determine. Whenever the tunes were adjusted, a short amount of time was needed before the loss rate stabilized. Sometimes it reacted quickly, while at other times it required a longer period before a specific loss rate could be determined. The number of protons in the test bunch was measured throughout the experiment period, and a calibration test was performed as mentioned in the previous section. This allowed the loss-rate data to be converted into lifetimes as shown in Fig.15.

The shaded scale shown on the right side of the scans indicates the lifetime, in hours, witnessed at each data point. In order to more effectively convey the regions of high and low lifetime, a two-dimensional interpolation algorithm turned the individual data points into a smooth, shaded surface. Contour lines are drawn at multiples of 20 hours.

Unfortunately, the regions covered by the two scans do not span exactly the same tune space, but there was a sufficient overlap to make the significant differences between the flattop and the Gaussian guns apparent. In Fig.15a, the flat-top gun usually produced poor lifetimes. This implies that the TEL-1 flattop gun tended to excite oscillations in at least some portion of the bunch particles, and the recorded lifetimes were mostly less than ten hours. However, in the tune space region near the main diagonal $Q_x = Q_y$, there is a relatively consistent pattern of lower losses. Along this strip, lifetimes as high as seventy hours were observed, almost as high as the lifetime of the bunch unaffected by the TEL-1.

The large regions of low lifetime again support the hypothesis that the flattop electron beam with sharp edges is adversely affecting protons. The outlying particles, witnessing strongly nonlinear focusing forces from the electron beam, do not survive as long as the core particles. Through the majority of the tested tune space, these particles are lost very quickly, and the gradual emittance growth of the core protons constantly feeds these losses. Only in a small working-point region do the outlying particles not escape so nimbly, slowing the rate at which particles are lost.

The second scan, in Fig.15b, shows the massive difference that the Gaussian gun had on the lifetime. The highest measured lifetimes were around 130 hours, almost indistinguishable from the bunch lifetime without the TEL-1. Much larger regions of lifetimes over twenty hours can also be seen. The fact that the highest lifetimes are nearly the same as for the unperturbed proton bunch bolsters the idea that TEL-1 fluctuations cannot, by themselves, remove particles from the bunch completely. Instead, we believe that the fluctuations contribute to a gradual emittance growth, and because there are no strongly nonlinear edges to the electron beam, the protons are still stable at larger orbits. This interpretation explains why a much larger percentage of the tested tune space offered moderate lifetimes than for the flat-top gun and why the best lifetimes observed are significantly longer with the Gaussian gun.

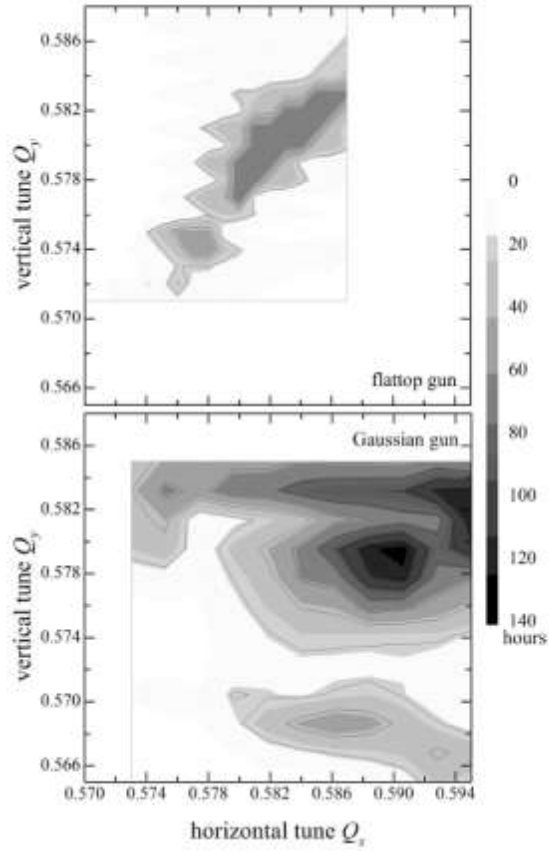

Figure 15 : Contour plots of proton bunch lifetime scans over a range of vertical and horizontal betatron tunes : a) top – with $dQ_x=0.004$ induced by TEL-1 with the flat-top electron gun ; b) bottom - with $dQ_x=0.004$ induced by TEL-1 with the Gaussian electron gun.

A significant improvement of the particle lifetime thanks to the employment of the Gaussian electron gun was critical for the first observation of a successful beam-beam compensation of the antiproton emittance growth.

## 5. COMPENSATION OF ANTIPROTON BEAM EMITTANCE GROWTH BY ELECTRON LENSES

After the installation of the Gaussian electron gun early in 2003 and the demonstration of a good proton lifetime with it, we successfully employed TEL-1 for beam-beam compensation in HEP stores – initially for the suppression of vertical emittance growth in the antiproton bunches.

In general, the beam-beam phenomena in the Tevatron collider are characterized by a complex mixture of long-range and head-on interaction effects, record high beam-beam parameters for both protons and antiprotons (the head-on tune shifts are about $\xi^p=0.020$ for protons and $\xi^a=0.028$ for antiprotons, in addition to long-range tune shifts of $\Delta Q^p=0.003$ and $\Delta Q^a=0.006$, respectively), and remarkable differences in beam dynamics of individual bunches. Figure 16 displays the Tevatron beam tunes at the beginning of a high-luminosity HEP store on a resonance plot. Particles with up to 6σ amplitudes are presented. Small amplitude particles have tunes near the tips of the "ties" depicted for all 36 proton and 36 antiproton bunches. The most detrimental effects occur when particle tunes approach the resonances. For example, an emittance growth of the core of the beam is observed near the fifth-order resonances (defined as $nQ_x+mQ_y=5$, such as $Q_{x,y}=3/5=0.6$) or fast halo particle loss near twelfth-order resonances (for example, $Q_{x,y}=7/12\approx 0.583$). Overall, the beam-beam effects at all stages of the Tevatron operation result in about 10-15% loss in a store's integrated luminosity for a well-tuned machine but it can often be as high as 20-30% in case of non-optimal operation [4].

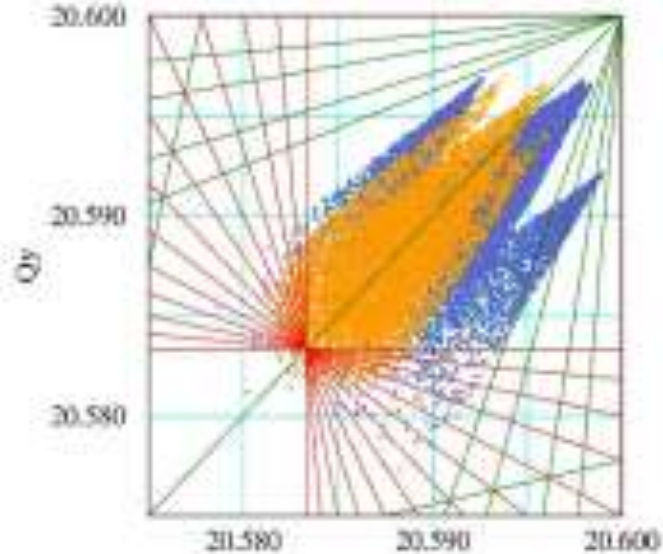

Figure 16 : Tevatron proton and antiproton tune distributions superimposed onto a resonance line plot. The red and green lines are various sum and difference tune resonances of up to twelfth order. The blue dots represent calculated the tune distributions for all 36 antiproton bunches; the yellow σ represent the protons. The tune spread for each bunch is calculated for particles up to 6σ amplitude taking into account the *measured* intensities and emittances.

In the Tevatron, the 36 bunches in each beam are arranged in three trains of 12 bunches each, and the spread of the intensities and emittances among the proton bunches is usually quite small. Consequently, a three-fold symmetry is expected [15] and observed [15, 4] in the pattern of the antiproton bunch orbits (which vary by about 40 microns bunch-to-bunch), tunes (which vary by up to 0.006 bunch-to-bunch) and chromaticities (6 units of $Q'=dQ/(dp/p)$ variations). It is not surprising that with such significant differences in orbits, tunes, and chromaticities, the antiproton bunch intensity lifetime and emittance growth rates vary considerably from bunch to bunch. As an illustration, Fig.17 shows the vertical emittance blowup early in an HEP store for all three trains of antiproton bunches.

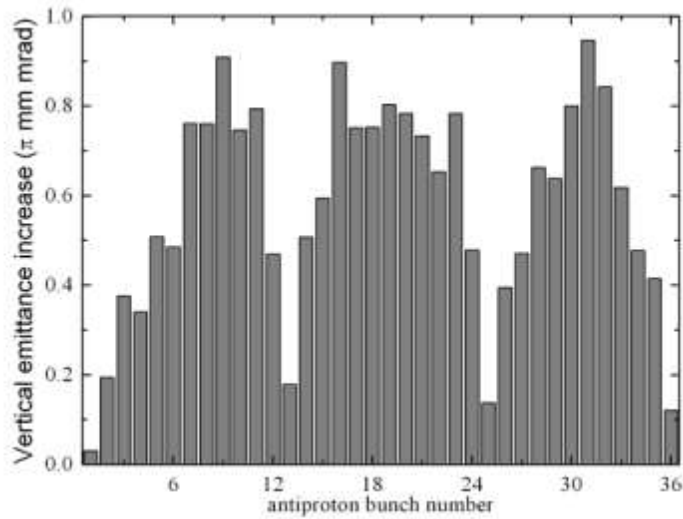

Figure 17 : Antiproton bunch emittance increase over the first 10 minutes after initiating collisions for HEP store #3231 with an initial luminosity $L=48\cdot 10^{30}\text{cm}^{-2}\text{s}^{-1}$.

One can see a remarkable distribution along the bunch train which gave rise to the term "scallops" (three "scallops" in three trains of 12 bunches) for this phenomenon – the end bunches of each train exhibit lower emittance growth than the bunches in the middle of the train. Because of the three-fold symmetry of the proton loading, the antiproton emittance growth rates are the same within 5-20% for corresponding bunches in different trains (in other words, bunches #1, #13, and #25 have similar emittance growths). The effect is dependent on the antiproton tunes, particularly on how close one bunch is to some important resonance. Typically, the Tevatron working points during 2003 were set to $Q_x = 0.582$ and $Q_y = 0.590$. At this working point, fifth-order (0.600), seventh-order (0.5714), and twelfth-order (0.583) resonances all play major roles in the antiproton beam dynamics. It was observed that vertical tune changes as small as -0.002 often resulted in a reduction of the amplitude of the "scallops." Smaller but still definite "scallops" were also seen in protons. After the initial 0.5-1 hour of each store, the growth rate of each bunch decreased significantly. This decrease is understood to be due to the steady decrease of the antiproton tune shift induced by the protons while the proton beam size grows and the proton intensity rapidly decreases at the beginning of a HEP store.

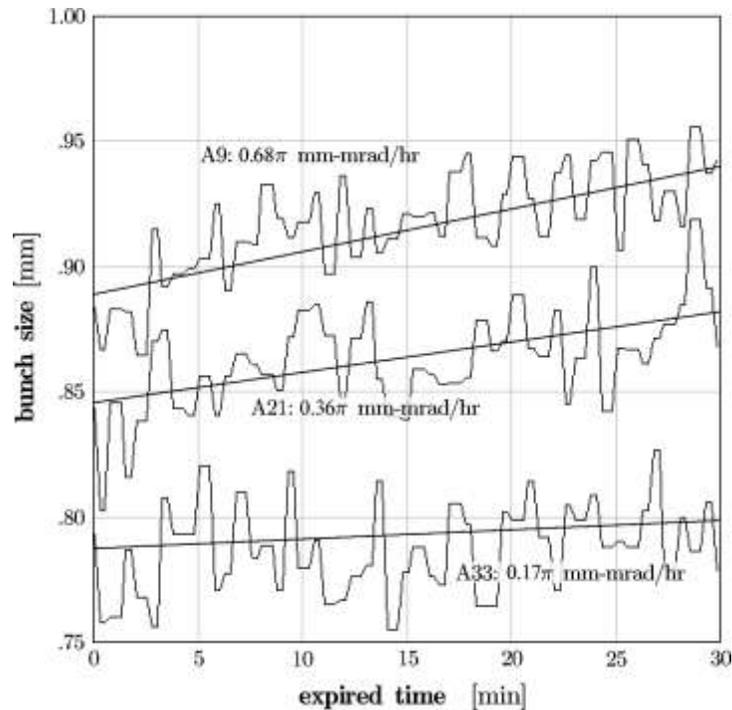

Figure 18. Evolution of antiproton bunch emittances over the fisrt 30 minutes of HEP store #2540: emittances for the ninth bunch in each of three trains are presented: #9, #21, and #33; TEL-1, with the Gaussian electron gun, is acting only on bunch #33.

TEL-1 was used at the beginning of several HEP stores in an attempt to reduce the emittance growth during the first half hour of collisions by acting on a single bunch. The goal was to significantly decrease the emittance growth of the particular bunch with respect to its "sibling" bunches (the equivalent bunches in the other two trains), or with respect to the same bunch in other, similar stores. TEL-1was timed on a single antiproton bunch at the beginning of the Tevatron stores and the vertical emittance growth of that antiproton bunch was monitored.

Figure 18 presents the evolution of the vertical rms sizes of three antiproton bunches (#9, #21, and #33) over the first 34 minutes after "initiating collisions" in store #2540 (May 13, 2003). The TEL was acting only on bunch #33. The bunch size was measured by a Synchrotron Light Monitor [9,16]. The corresponding emittance growth was $0.68 \pm 0.06\pi$ mm-mrad/hr for bunch #9, $0.37 \pm 0.07\pi$ mm-mrad/hr for #21, but only $0.16 \pm 0.07\pi$ mm-mrad/hr for #33, shown by the fitted lines in Fig.18. During this experiment, TEL-1 was set to a current of 0.6 A, an energy of 4.5kV, and an rms beam size of 0.8 mm. Given an interaction length of 2.0 m, the expected maximum horizontal antiproton tune shift is -0.004, and the expected vertical one is -0.001. After 34 minutes the TEL1 was turned off, and the emittances of all three bunches leveled.

Several attempts were made to test this ability, each on a new store during the first short period. These tests, along with pertinent parameters, are summarized in Table 2, where each row represents a different store (the designated store number is listed down the left column).

Table 2: Growth of the rms vertical antiptoton bunch emittance in the beginning of stores. All of the growth numbers are in units of π mm-mrad/hr, with a typical fit error of ±0.07 π mm mrad/hr. For the indicated stores, TEL-1 was acting on bunch #33.

| Store # | duration | #9 growth | #21 growth | #33 growth | |
|---------|----------|-----------|------------|------------|----------|
| #2536   | 40 min   | 1.65      | 1.53       | 1.55       |          |
| #2538   | 35 min   | 0.32      | 0.28       | 0.46       |          |
| #2540   | 34 min   | 0.68      | 0.37       | 0.17       | TEL-1 on |
| #2546   | 30 min   | 0.65      | 0.32       | 0.67       | TEL-1 on |
| #2549   | 26 min   | 0.75      | 0.60       | 1.18       | TEL-1 on |
| #2551   | 34 min   | 1.12      | 1.1        | 1.17       |          |

In the three stores listed without TEL-1 in Table 2, the emittance growth rate of bunch #33 is similar or just slightly larger than that of its siblings. In stores #2546 and #2549, it is still larger. However, in store #2540, the growth of bunch #33 is significantly less than that of the other two bunches. The differences between consecutive stores are considerable, but the only intentional difference is the application of TEL-1. Soon after store #2551, a set of sextupole correction magnets was employed to lower the antiproton tune sufficiently enough (without affecting proton tunes) so that the scallops were avoided and there was no operational need in usage of TEL-1 for that purpose anymore.

The effect of TEL-1 in stores #2540, #2546 and #2549 is obvious, though not well-controlled, since it can have an adverse effect instead (store #2549, for example). We believe that the uncertainty is due to insufficiently precise centering of the electron beam on the antiproton orbit. There were two reasons for that: the antiproton orbit itself changes from store to store by as much as 1 mm at the TEL-1 location; and the BPMs used in TEL-1 have an observable 0.5–1.5 mm systematic difference between the nanosecond-scale antiproton bunch and the microsecond-scale electron pulse scales (though the statistical accuracy of either measurement was about 30 microns). Such a large error in the BPM measurement led to difficulties in the experiment repeatability.

## 6. IMPROVEMENT OF PROTON BEAM INTENSITY LIFETIME AND LUMINOSITY LIFETIME BY ELECTRON LENSES

In 2004-2006, we have introduced four very important changes in the Tevatron and TEL operation which allowed a very regular, repeatable and successful employment of electron lenses for beam-beam compensation. Firstly, the Tevatron automatic orbit stabilization system was installed and commissioned [17], so that typical store-to-store orbit changes as well as low-frequency orbit drifts during HEP stores do not exceed 0.1 mm (high frequency orbit jitter is still uncorrected, but it is not very significant - about 0.02-0.04 mm peak-to-peak). Secondly, a new signal processing technique was introduced for the TEL BPMs which reduced the frequency dependence of the monitors from 0.5-1.5 mm down to about 0.1 mm [18].

Thirdly, the second electron lens was built and installed at the A11 location of the Tevatron ring [19, 8] that allowed us to conduct dedicated beam-beam compensation studies very often – ultimately, in every HEP store – with one of the lenses while the other one is always dedicated to the abort gap cleaning (a standard TEL job since early Run II – see Refs. [20, 16]). Last, but not least, electron guns with "smoothed-edge-flat-top" current distributions were designed, built and installed in both TEL-1 and TEL-2 [6]. All the results presented in this Chapter are obtained with the SEFT electron guns. After commissioning of all these features and attainment of stable operation, a significant improvement of proton beam lifetime under the action of electron lenses was demonstrated [1].

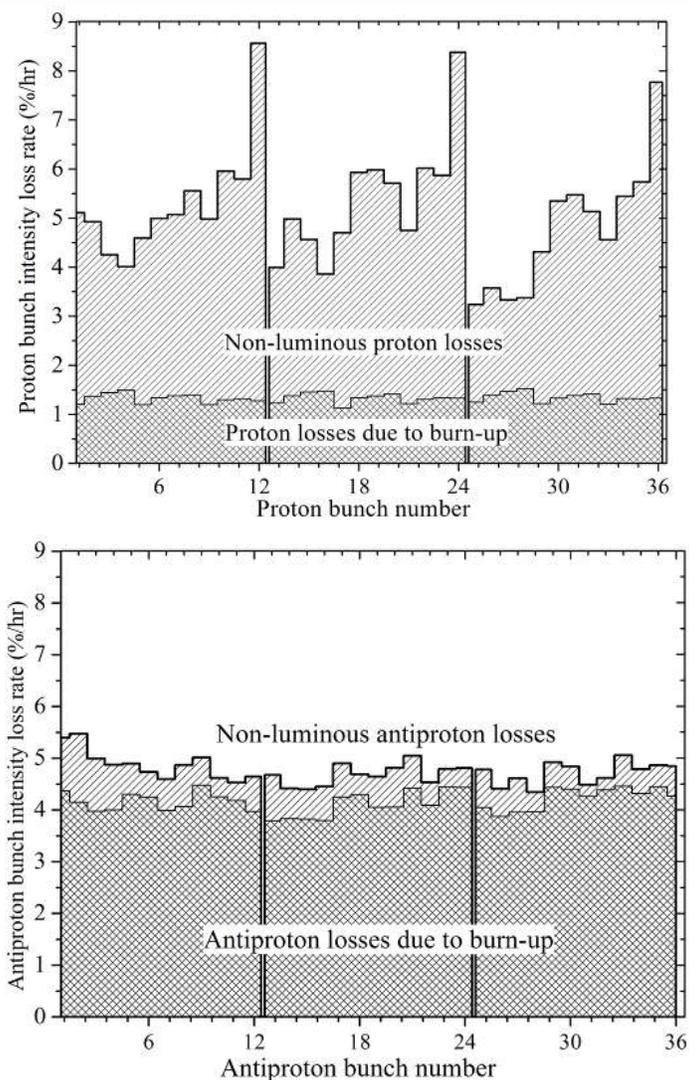

Figure 19: a) top - proton-bunch intensity loss rates and b) bottom - antiproton-bunch intensity loss rates at the beginning of the Tevatron store #5155, Dec. 30, 2006, with an initial luminosity $L=250\times10^{30}$ cm$^{-2}$s$^{-1}$ .

A significant attrition rate of the protons due to their interaction with the antiproton bunches, both in the main IPs and in the numerous long-range interaction regions is one of the most detrimental effects of the beam-beam interaction in the Tevatron [4]. This effect is especially large at the beginning of the HEP stores where the total proton beam-beam tune shift induced by the antiprotons at the two main IPs (B0 and D0) can reach the values of $2\xi^{proton}=+0.020$. Fig. 19a) shows a typical distribution of proton loss rates at the beginning of a high-luminosity HEP store. Bunches #12, 24, and 36 at the end of each bunch train typically lose about 9% of their intensity per hour while other bunches lose only 4% to 6% per hour. These losses are a very significant part of the total luminosity decay rate of about 20% per hour (again, at the beginning of the high luminosity HEP stores). The losses due to inelastic proton-antiproton interactions $dN_p/dt=\sigma_{int} L$ at the two main IPs ($\sigma_{int}=0.07$ barn) are small (1–1.5%/hr) compared to the total losses. Losses due to inelastic interaction with the residual vacuum are less than 0.3%/hr. The single largest source of proton losses is the beam-beam interaction with the antiprotons. Such conclusion is also supported by Fig.19 a), which shows a large bunch-to-bunch variation in the proton loss rates within each bunch train, but very similar rates for equivalent bunches, e.g. bunches #12, 24, and 36. On the contrary, antiproton intensity losses $dN_a/dt$ are about the same for all the bunches – see Fig. 19 b) – as they are mostly due to luminosity burn-up and not determined by beam-beam effects.

The remarkable distribution of the proton losses seen in Fig.19, e.g. particularly high loss rates for bunches #12, 24, 36, is usually thought to be linked to the distribution of betatron frequencies along the bunch trains bunch. Bunches at the end of the trains have their vertical tunes closer to the 7/12≈0.583 resonance lines – see Fig.20 – and, therefore, the higher losses. The average Tevatron proton tune $Q_y$ of about 0.*588-0.589* lies just above this resonance, and the bunches at the end of each train, whose vertical tunes are lower by $\Delta Q_y=-(0.002-0.003)$ due to the unique pattern of long-range interactions, are subject to stronger beam-beam effects [4]. The tunes $Q_y$ $Q_x$ are carefully optimized by the operation crew to minimize the overall losses of intensity and luminosity. For example, an increase of the average vertical tune by quadrupole correctors is not possible because it usually results in higher losses and "scallops" as small amplitude particle tunes move dangerously close to the 3/5=0.600 resonance (see Fig.16).

When properly aligned, the TEL-2 electron beam focuses protons and, thus, produces a positive vertical tune shift of the proton bunch it acts on, proportional to the electron current – as depicted in Fig. 5 – and, therefore, it should reduce the losses. A preliminary alignment of the electron beam has been done by relying on the TEL beam position measurement system. However, additional fine tuning is usually necessary to achieve the best possible compensation. Measurements of the proton loss rate versus the electron beam position with an increased electron current were performed at the very end of a store, when no beam-beam related losses occur. This approach allowed us to determine the optimal electron beam position. Since the Tevatron orbit is kept stable by the orbit feedback system within 100 μm, the end-of-store values can be used throughout other stores, unless an optics change is introduced.

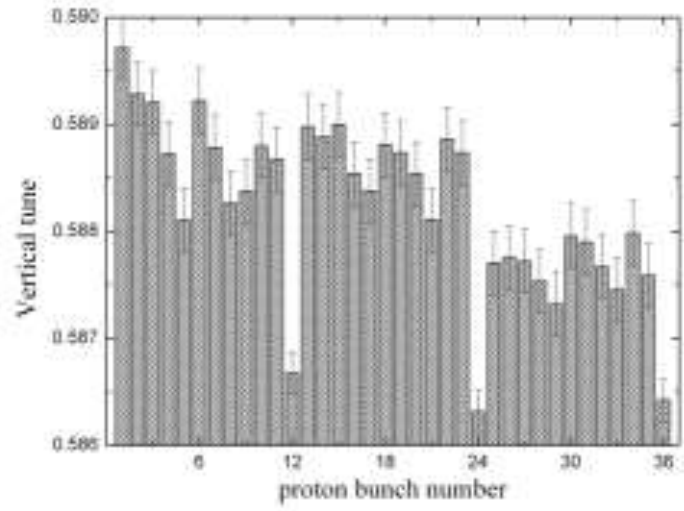

Figure 20 : Proton bunch tunes measured by Digital Tune Monitor [21] at the beginning of store #5301 (March 3, 2007) with an initial luminosity $L=203\times10^{30}$ cm$^{-2}$s$^{-1}$

In one of the very first BBC demonstration experiment, we timed the TEL-2 electron pulse onto bunch P12 without affecting the other bunches. The transverse alignment of the electron beam is illustrated in Fig.3 above. Figure 21a) shows that when the TEL peak current was increased to 0.3A, the lifetime $\tau=N_p/(dN_p/dt)$ of bunch P12 went up to 17.4±0.1 hours from 8.75±0.1 hours (in other words, the loss rate $1/\tau$ improved two-fold from about 11.4%/hr to some 5.7%/hr). At the same time, the lifetime of bunch P24, an equivalent bunch in another bunch train, remained low and did not change significantly ($\tau$=8.66 hour lifetime slightly improved to 10 hours due to natural reasons – see discussion below). The TEL was left on P12 for the first 1.5 hours of the store and the intensity decay of that particular bunch was one of the lowest among all 36 proton bunches – as shown in Fig.21 b) which presents the loss rates corrected for luminous intensity decay which is not due to beam-beam effects $(dN_p/N_p)_{NL}/dt=(dN_p/N_p)_{total}/dt - \sigma_{int} L/N_p$. It is noteworthy, that the vertical tune shift caused by such a moderate electron current $J_e$ =0.3A is about $dQ_y$=+0.0007 (as can be seen in Fig.5b) and it is not sufficient for P12 to reach the average tune $dQ_y < |\Delta Q_y| \approx 0.002$. Therefore, the TEL-induced tune shift could not be considered as the only mechanism responsible for the significant lifetime improvement in that experiment.

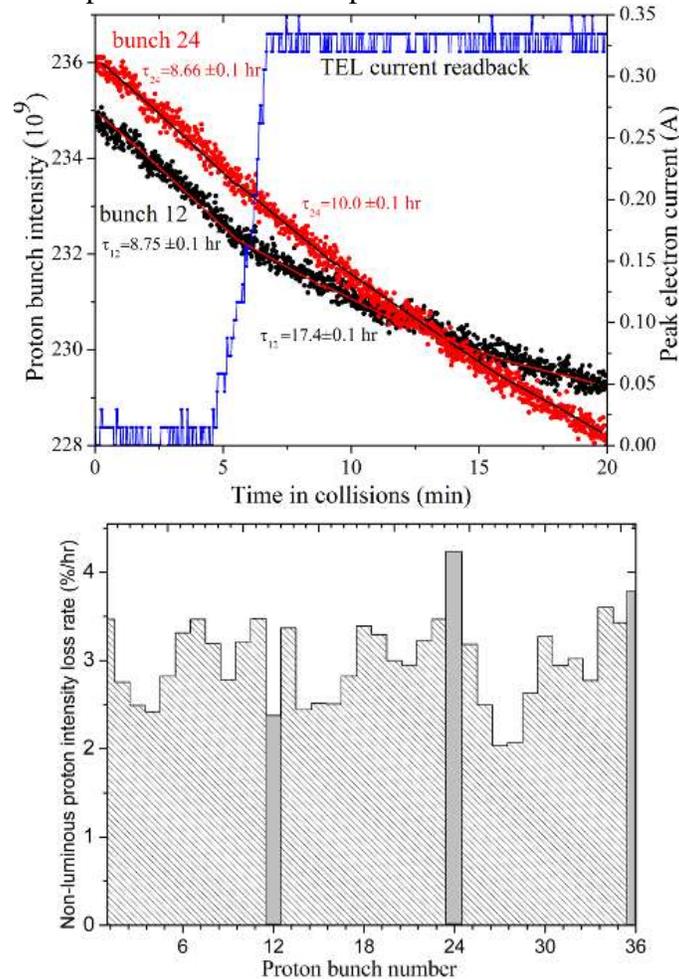

Figure 21: a) Intensity decay of proton bunch 12 affected by the TEL-2 and reference bunch 24 at the beginning of store #5123 with an initial luminosity $L$=197×10$^{30}$ cm$^{-2}$s$^{-1}$. The blue line shows the measured TEL-2 peak current; b) the average non-luminous bunch intensity loss rate in the first 1.5 hours of the store.

To explore the electron lens effects in more detail, another series of BBC studies has been performed in one of the highest luminosity Tevatron stores #5183, in which the TEL-2 operated in a DC regime with $J_e$=0.3A – thus, providing the same effect on all proton bunches in the beam – and has been regularly turned off and on. When the TEL-2 was turned on at the very beginning of the store, it improved the intensity lifetime of all the bunches, as presented in Fig.22, although the largest improvement $R$=2.2, defined as the ratio of the proton lifetime with the TEL and without it, has been observed for bunches P12, P24 and P36, as expected. Later in the store, the TEL-2 has been turned off for some 20 minutes, then, by use of magnet correctors, an equivalent tune change of $dQ_y \approx$0.0008 has been introduced and the beam intensity decay measured for some 20-30 minutes. After that, the tune correction has been turned off for 20-30 minutes for "reference" lifetime measurement, which was followed by another ½ hour of TEL-2 operation, and so on. Figure 23 shows the total proton intensity lifetime measured for each of these intervals. One can see that initially the beam lifetime improves every time when either TEL2 or the tune correction has been introduced. Nevertheless, after some 5 hours into the store, the TEL-2 still lead to the lifetime improvement while during two periods of tune correction the lifetime slightly decreased with respect to the unperturbed reference periods (the black bars in Fig.23).

Besides a significant reduction of the proton intensity loss rates, the luminosity lifetime $\tau_L$=$L/(dL/dt)$ has been improved as well. Figure 24a) compares the changes of the lifetime of the combined luminosity for the three bunches #12,24 and 36 due to TEL-2 and due to the tune correction in the same store #5183. The height of each bar is equal to :

$$R_L = \frac{2\tau_L(with\ TEL\ or\ dQ_y\ change)}{\tau_L(reference\ period\ before\ the\ change) + \tau_L(reference\ period\ after\ the\ change)}. \quad (4)$$

The luminosity lifetime improvement due to TEL-2 is about 12% at the beginning of the store. Later in the store the TEL-2 effect was somewhat larger than that of the global tune correction $dQ_y$. The evolution of the average proton and antiproton tunes is shown in Fig.24b).

The TEL induced improvements in the luminosity lifetime of about 10% are significantly smaller than the corresponding changes in the proton intensity lifetime (about a factor of 2) because the luminosity decay is driven mostly by other factors, the strongest being the proton and antiproton emittance increase due to intra-beam scattering and the antiproton intensity decay due to luminosity burn-off. Usually, these factors combined lead to the decay of instantaneous luminosity approximately given by [4]

$$L(t) = \frac{L_0}{1 + t/\tau_L} \quad (5)$$

so that the total integrated luminosity over a store is proportional to the product of the initial luminosity and the luminosity lifetime $L_0 \tau_L ln(1+ T/\tau_L)$, where $T$ denotes the store duration. Therefore, a 10% improvement of the luminosity lifetime $\tau_L$ due to TEL-2 results in a proportional increase of the integrated luminosity.

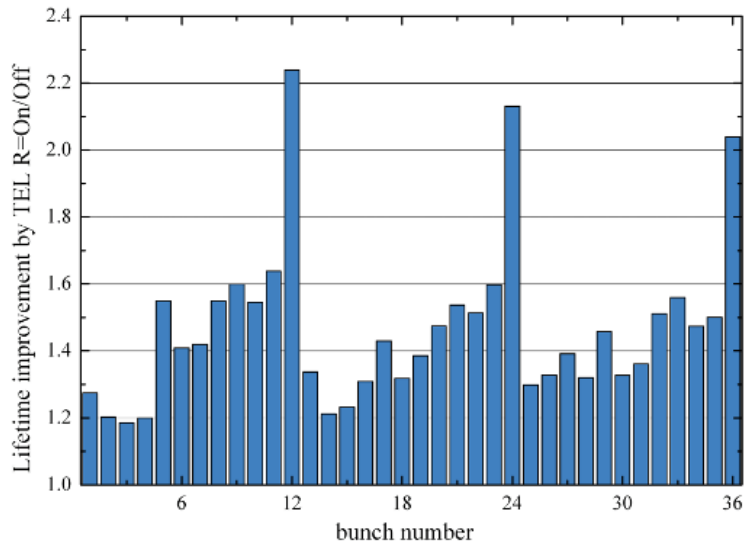

Figure 22: Proton bunch lifetime improvement due to TEL-2 (DC regime) early in store #5183 with initial luminosity $L=253\times10^{30}$ cm$^{-2}$s$^{-1}$.

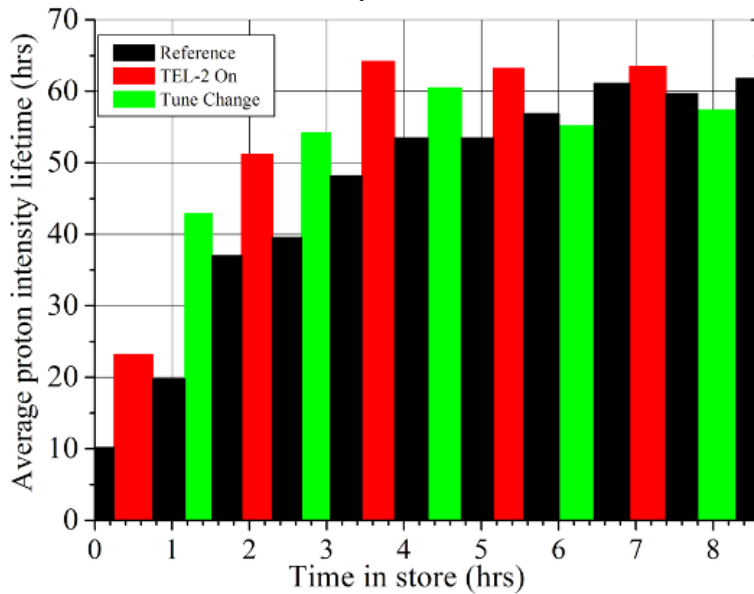

Figure 23: Average proton intensity lifetime $dt/(dN_p/N_p)_{total}$ in store #5183 when, repetitively, the TEL-2 was turned on protons with 0.3A of DC electron current (red bars), then turned off for reference (black bars), next the proton vertical tune was shifted up 0.0008 by quadrupole and sextupole tune correctors (green bars), and the correctors were finally turned off again (black).

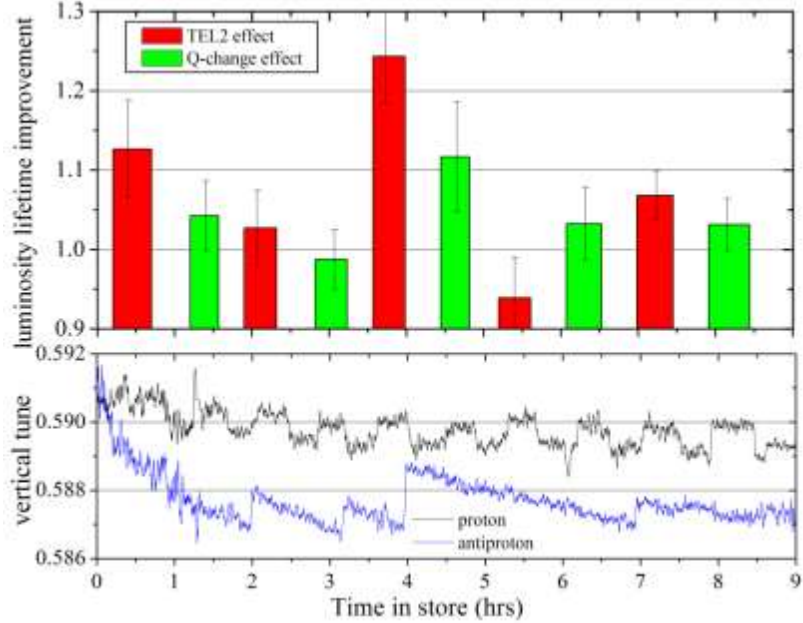

Figure 24: a) top- Luminosity lifetime improvement by TEL-2 and by tune correctors in store #5183; b) bottom – the average vertical tunes of the proton and antiproton bunches measured by the 1.7 GHz Schottky detectors.

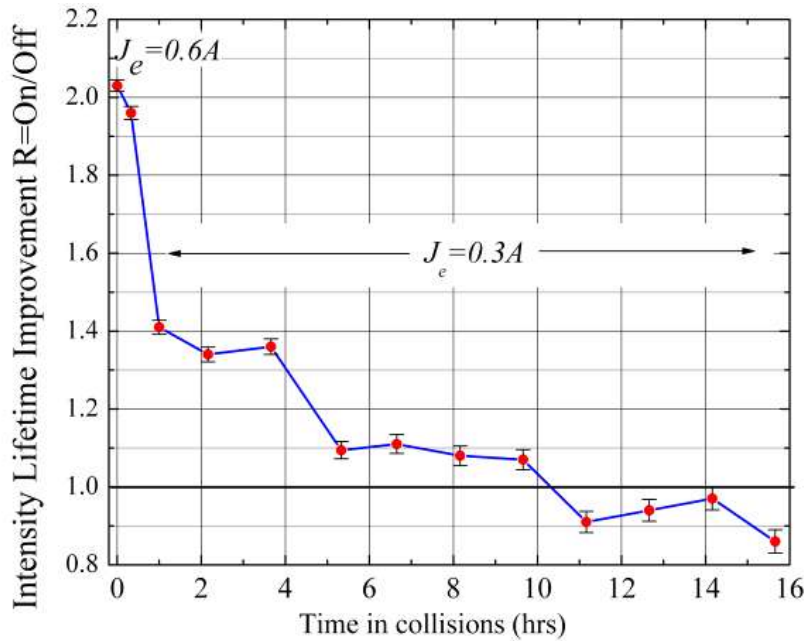

Figure 25: Relative improvement of the TEL induced proton bunch #12 lifetime vs. time (store #5119, Dec. 12, 2006, initial luminosity $L=159\times10^{30}$ cm$^{-2}$s$^{-1}$)

Usually, the proton lifetime, dominated by beam-beam effects, gradually improves with time in a HEP store and reaches some 50-100 hours after 6-8 hours of collisions. This is due to the decrease in the antiproton population and to an increase in antiproton emittance, both contributing to a reduction of the proton beam-beam parameter $\xi^p$. In store #5119, we studied the effectiveness of the beam-beam compensation by repeatedly turning TEL-2 on a single bunch P12 and off every half hour for 16 hours. The relative bunch intensity lifetime improvement *R* is plotted in Fig.25 [1]. The first two data points correspond to $J_e$=0.6A , but subsequent points were taken with $J_e$=0.3A to observe the dependence of the compensation effect on the electron current. The change of the current resulted in a drop of the relative improvement from *R*=2.03 to *R*=1.4. A gradual decrease in the relative lifetime improvement is visible until after about ten hours, where the ratio reaches 1.0 (i.e., no gain in the lifetime). At this point, the beam-beam effects have become very small, providing little to compensate. Similar experiments in several other stores with initial luminosities ranging from $1.5 \cdot 10^{32}$ cm$^{-2}$ s$^{-1}$ to $2.5 \cdot 10^{32}$ cm$^{-2}$ s$^{-1}$ reproduced these results.

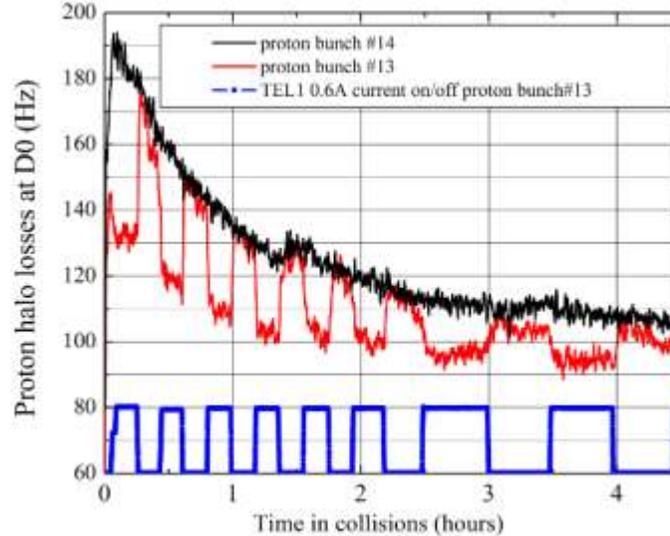

Figure 26: Proton bunch intensity loss rates detected by D0 counters: black – for reference bunch 14, red – for bunch 13 affected by TEL-1 (first 4 hours in store #5352 *L*=197×10$^{30}$ cm$^{-2}$s$^{-1}$)

Comparable improvement of the proton intensity lifetime (up to 40%) has been observed in experiments performed with TEL-1. The only design difference between the two lenses is that the TEL-1 bending section has a 90 degree angle between the gun solenoid and the main solenoid while this angle is about 57 degrees for TEL-2 (depicted in Fig.1). TEL-1 is installed in a location with large horizontal beta-function and mostly shifts horizontal proton tune up. As the proton horizontal tunes are lower by $\Delta Q_x \approx (0.002-0.003)$ for the bunches at the beginning of the bunch trains, P1, P13, and P25 [4], the TEL-1 effect is the largest for them. The reduction of the global proton loss rates due to the electron lenses can easily be seen by the local halo loss rate detectors installed in the D0 and CDF detectors, which can measure the losses on a bunch-by-bunch basis. Figure 26 shows the dependence of D0 proton loss rate on the TEL-1 electron current. In this experiment TEL-1, being a horizontal beam-beam compensation device, was acting on P13 which has the lowest horizontal tune. Bunch P14 – unaffected by TEL-1 - was chosen as a reference bunch because its behavior in terms of halo and lifetime was very similar to P13, without TEL. The loss rate of P13 dropped by about 35% once a 0.6A-peak electron current was turned on, while the P14 loss rate stayed unaffected. After about 12 min the e-current was turned off which made the P13 loss rate return to the reference level. The loss reduction has been repeated several times over the next 4 hours in this store and it was confirmed in several other HEP stores.
.

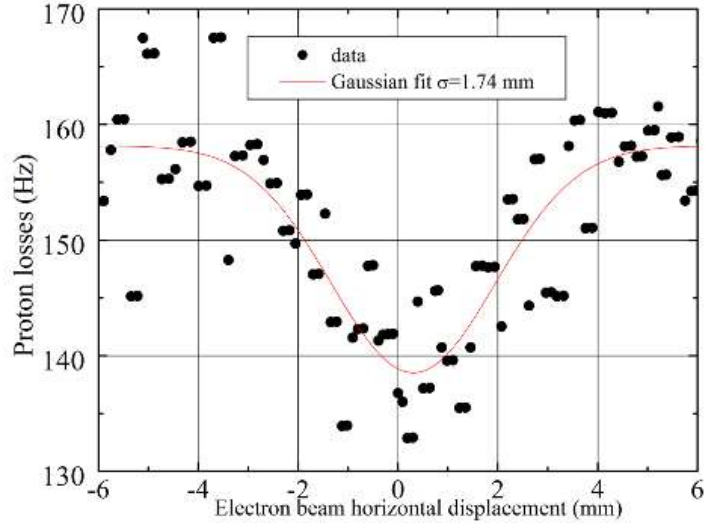

Figure 27: Proton loss rate measured at the D0 detector vs horizontal displacement of electron beam in TEL-1.

We have also studied the dependence of the loss-rate reduction on the electron beam position with respect to the proton beam position. Figure 27 demonstrates that if the electron beam is displaced from the proton orbit by more than 4 mm, the effect of the TEL-1 on the D0 proton halo loss rate vanishes.

## 7.     DISCUSSION AND CONCLUSIONS

Superconducting hadron colliders - Tevatron, HERA, RHIC and LHC, being the most notable examples - are the most sophisticated instruments for the cutting edge high-energy physics research. The effectiveness and luminosity of such complex and costly machines is limited by the EM beam-beam interaction. Contrary to electron-positron storage rings, the beam-beam effects in hadron colliders do not set a hard limit on performance. Instead they lead to bigger and bigger losses or emittance blow-ups until they result in either intolerable energy deposition in the superconducting magnets (quenches) or in intolerable detector halo backgrounds. Due to the absence of synchrotron-radiation damping, numerical simulations of beam-beam effects in hadron colliders are very cumbersome, extremely time consuming and have not yet reached the level of satisfactory trustworthiness. Therefore, experimental studies of beam-beam effects and especially, their compensation are of utmost importance. All reasonable beam-beam compensation proposals become the subject of detailed experimental studies. The outcome of studies on four-beam compensation, the use of octupole magnets for compensating beam-beam nonlinearities, and the long-range beam-beam compensation with use of current-carrying wires is overviewed in Ref.[1] and references therein. Beam-beam compensation with low energy electron beams has many advantages compared to the afore-mentioned schemes. This article, together with Ref.[1], summarizes the results of the pioneering studies with the Tevatron electron lenses.

Major outcomes of our work are the development of electron lenses, the demonstration of their compatibility with the operation of superconducting hadron collider and the experimental proof of compensation of beam-beam effects in the Tevatron proton-antiproton collider.

The results of the BBC studies presented in this article demonstrate that the TELs shift the proton and antiproton tunes as originally predicted in Ref. [5] and Eq.(3). Both lenses built – TEL-

1 and TEL-2 - produce a very strong positive effect on the lifetime of the Tevatron proton bunches which otherwise suffered most from the collisions with antiprotons. The observed lifetime improvement at the beginning of a HEP store (when the beam brightness and luminosity are highest, and the beam-beam interaction is strongest) can be as big as a factor of 2. Only some 10 hours into the stores, the beam-beam effects and the BBC gains decrease to insignificant levels. The beam-beam compensation effect is found to be tune dependent and somewhat outperforming the traditional tune correction method. It has to be noted that the difference between two electron lenses - bending angle of the electron trajectory is 90 degrees in one lens and 57 degrees in another – did not significantly impact the reduction of the proton losses by both lenses.

We experimentally learned that for the successful operation of electron lenses one needs a smooth transverse distribution of the electron current density, a good alignment of the electron electron beam on the beam of interest – within a fraction of proton or antiproton rms beam size; and low noises and ripples in the electron beam current and position.

We have observed the reduction of the antiproton emittance growth rate in some early BBC studies with TEL-1, but the effect was not reproduced reliably, because of poor control of the electron beam centering on the antiprotons.

We have not seen any sign of coherent instabilities due to the (anti) proton beam interaction with the electron beam, despite initial concerns [5].

Naturally, as the next step of the BBC program, we plan to incorporate the Tevatron Electron Lenses into the routine operation of the Tevatron collider.

The versatility of the electron lenses allows their use for many other purposes, e.g. for the removal of unwanted DC beam particles leaking out of RF buckets into the Tevatron abort gaps between the bunch trains [16]. There are also proposals to use electron lenses for space-charge compensation in high intensity proton synchrotrons [22], for the reduction of a tune spread in proton-proton or like-charge colliding beams [23, 5], and for beam collimation in the LHC [24]. An LHC electron lens with about 2.4 A of DC current can compensate head-on effects induced by collisions with 2.3e11 proton bunches, twice the LHC nominal bunch intensity[25]. As such, the electron lenses combined with current carrying wires for long-range beam-beam compensation are believed to allow reaching higher collider luminosities without a significant increase of particle loss rates or emittance growth rates.

**ACKNOWLEDGEMENTS**

We would like to thank J.Annala, T.Bolshakov, A.Burov, R.Dixon, B.Drendel, J.Featherstone, D.Finley, R.Hively, S.Holmes, A.Jansson, A.Klebaner, M.Kufer, V.Lebedev, J.Marriner, A. Martinez, S.McCormick, E.McCrory, D.McGinnis, N.Mokhov, R.Moore, J.Morgan, M.Olson, H.Pfeffer, D.Wolff, V.Scarpine, G.Saewert A.Shemyakin, J.Steimel, A.Valishev (FNAL), A.Kuzmin, M. Tiunov (BINP, Novosibirsk), S.Kozub, V. Sytnik, L. Tkachenko (IHEP, Protvino), V.Danilov (ORNL), W.Fischer, Y.Luo, S.Peggs (BNL), E.Tsyganov (South-Western Medical Center), and A.Seryi (SLAC) for valuable technical contributions, assistance during beam studies and useful discussions on the subject. Fermilab is operated by Fermi Research Alliance, LLC under Contract No. DE-AC02-07CH11359 with the United States Department of Energy.